\documentclass[twocolumn,english,prl,amssymb,aps,showpacs,twocolumn,amsmath,showkeys,floatfix]{revtex4-1}
\usepackage[T1]{fontenc}
\usepackage[latin9]{inputenc}
\usepackage{geometry}
\usepackage[active]{srcltx}
\usepackage{color}
\usepackage{amstext}
\usepackage{graphicx}
\usepackage{esint}

\makeatletter

\@ifundefined{textcolor}{}
{%
 \definecolor{BLACK}{gray}{0}
 \definecolor{WHITE}{gray}{1}
 \definecolor{RED}{rgb}{1,0,0}
 \definecolor{GREEN}{rgb}{0,1,0}
 \definecolor{BLUE}{rgb}{0,0,1}
 \definecolor{CYAN}{cmyk}{1,0,0,0}
 \definecolor{MAGENTA}{cmyk}{0,1,0,0}
 \definecolor{YELLOW}{cmyk}{0,0,1,0}
}

\usepackage{babel}

\makeatother

\usepackage{babel}
\begin{document}

\title{Inelastic photon scattering via the intracavity Rydberg blockade}

\author{A. Grankin$^{1}$, E. Brion$^{2}$, R. Boddeda$^{1}$, S. \'{C}uk$^{1}$,
I. Usmani$^{1}$, A. Ourjoumtsev$^{1,\dagger}$, P. Grangier$^{1}$ }

\affiliation{$^{1}$Laboratoire Charles Fabry, Institut d'Optique Graduate School,
CNRS, Université Paris-Saclay, 91127 Palaiseau, France, \\
 $^{2}$Laboratoire Aimé Cotton, Université Paris-Sud, ENS Cachan,
CNRS, Université Paris-Saclay, 91405 Orsay Cedex, France.}
\begin{abstract}
Electromagnetically induced transparency (EIT) in a ladder system
involving a Rydberg level is known to yield giant optical nonlinearities
for the probe field, even in the few-photon regime. This enhancement
is due to the strong dipole-dipole interactions between Rydberg atoms
and the resulting excitation blockade phenomenon. In order to study
such highly correlated media, ad hoc models or low-excitation assumptions
are generally used to tackle their dynamical response to optical fields.
Here, we study the behaviour of a cavity Rydberg-EIT setup in the
non-equilibrium quantum field formalism, and we obtain analytic expressions
for elastic and inelastic components of the cavity transmission spectrum,
valid up to higher excitation numbers than previously achieved. This
allows us to identify and interpret a polaritonic resonance structure,
to our knowledge\textsl{ }\textsl{\emph{unreported }}so far. 
\end{abstract}


\maketitle

\paragraph*{Introduction.}

Optical quantum information processing requires photonic gates. For
the sake of efficiency it is preferable to implement them in a deterministic
way, which implies photon-photon interactions. Though impossible to
achieve directly, such interactions can be effectively emulated by
coupling photons to an atomic ensemble driven in a Rydberg-EIT configuration
either in free-space \cite{GNP2013,GOFP2013} or in a cavity setup
\cite{SPB13,PBS12}. The full dynamics of such a strongly correlated
many-body system and therefore its effects on the incoming photons
cannot be computed exactly. So far, analytic expressions of dynamical
variables like, e.g., the correlation functions of the transmitted
field could be derived either using \emph{ad hoc} models -- such as
the Rydberg bubble picture \cite{GNP2013,GBB14}, or resorting to
the perturbation theory restricted to the lowest non-vanishing order
in the number of incoming photons \cite{GOFP2013,GBB15}. In this
Letter, we employ the Schwinger-Keldysh contour formalism \cite{S61,R07,SL13}
to derive analytic expressions for field correlation functions in
a cavity setup beyond the lowest non-vanishing order in the excitation
number \cite{GBB15}. By opening a systematic and manageable way to
deal with higher-order terms, our approach breaks new ground for solving
the outstanding problem set by the many-body dynamics of Rydberg-blockaded
ensembles interacting with quantized light. It also allows us to unveil
nontrivial physical features of the transmitted light spectrum that
we explain by a simple polaritonic picture.

We consider an ensemble of $N$ atoms with a ground, intermediate
and Rydberg states, denoted by $\left\vert g\right\rangle $, $\left\vert e\right\rangle $
and $\left\vert r\right\rangle $, respectively, loaded in an optical
cavity \cite{GBB14}. The transitions $g\leftrightarrow e$ and $e\leftrightarrow r$
are driven by the cavity mode, of frequency $\omega_{c}$ and annihilation
operator $a$, and the strong control field, with the coupling strength
$g$ and the Rabi frequency $\Omega_{cf}$, respectively. The cavity
is fed through an input mirror with decay rate $\gamma_{c}^{\left(f\right)}$
by a weak probe laser of frequency $\omega_{p}$, while the field
transmitted by the cavity can be detected through an output mirror
with decay rate $\gamma_{c}^{\left(d\right)}$; we moreover set $\gamma_{c}\equiv\gamma_{c}^{\left(f\right)}+\gamma_{c}^{\left(d\right)}$.
We define detunings for the cavity $\Delta_{c}=\left(\omega_{p}-\omega_{c}\right)$,
single-photon $\Delta_{e}=\left(\omega_{p}-\omega_{eg}\right)$ and
two-photon $\Delta_{r}=\left(\omega_{p}+\omega_{cf}-\omega_{rg}\right)$,
with respect to the frequencies $\omega_{eg}$ and $\omega_{rg}$
of the $g\leftrightarrow e$ and $g\leftrightarrow r$ transitions.
We denote by $\gamma_{e}$ and $\gamma_{r}$ the decay rates from
the intermediate $\left\vert e\right\rangle $ and Rydberg $\left\vert r\right\rangle $
states, respectively. 

If there were no atomic interactions, the cloud driven under perfect
EIT conditions $\left(\gamma_{r}\approx\Delta_{r}\approx0\right)$,
would be transparent for the probe light. The dipole-dipole-interaction-induced
blockade phenomenon \cite{SWM10,LFC01} actually prevents most of
the atoms in the sample from being Rydberg excited. If $\Delta_{e}\approx0$,
spontaneous emission from the intermediate state is strongly enhanced
which significantly modifies the shape of the transmitted light spectrum.
This effect can be characterized by the steady state correlation function
of the intracavity light $\left\langle a^{\dagger}\left(t\right)a\left(0\right)\right\rangle $
\cite{WM08}. At the lowest non-vanishing order in the feeding rate
$\left|\alpha\right|\equiv\sqrt{2\gamma_{c}^{\left(f\right)}I_{in}}$,
where $I_{in}$ is the incident photon flux fed into the cavity, the
correlation function was shown to factorize, \emph{i.e.} $\left\langle a^{\dagger}\left(t\right)a\left(0\right)\right\rangle ^{\left(2\right)}=\left\langle a^{\dagger}\left(t\right)\right\rangle ^{\left(1\right)}\left\langle a\left(0\right)\right\rangle ^{\left(1\right)}$
\cite{GBB15}, where the superscript denotes the order in $\alpha$.
To reveal nonlinear features, one has to investigate orders higher
than four -- by conservation of excitation number the third order
vanishes. Usual techniques are not suited to this task. In particular,
the standard fourth-order perturbative expansion would already lead
to a cumbersome hierarchy of Heisenberg equations which could hardly
be generalized further. Here, we show the Schwinger-Keldysh contour
formalism \cite{S61,R07,SL13} allows one to compute dynamical variables
of the system up to \emph{a priori} arbitrary order in the feeding
strength, in a systematic and handy way. Besides bringing physical
insight into the specific problem considered here, our calculation
demonstrates how powerful this approach is to deal with non-equilibrium
dynamics of atomic systems as already stressed in \cite{FY99}.

\paragraph*{Dynamical equations of the system. }

According to Holstein-Primakoff approximation, the atomic lowering
operators $\sigma_{ge}^{\left(n\right)}$ and $\sigma_{gr}^{\left(n\right)}$
can be treated as bosons $b_{n}$ and $c_{n}$, respectively, in the
low excitation regime \cite{GBB15,BCF14}. Performing the rotating
wave and Markov approximations, the relevant Heisenberg-Langevin equations
are 
\begin{eqnarray}
\frac{d}{dt}a & = & \mbox{i}\left[H,a\right]-\gamma_{c}a+\sqrt{2\gamma_{c}^{\left(f\right)}}a_{in}^{\left(f\right)}+\sqrt{2\gamma_{c}^{\left(d\right)}}a_{in}^{\left(d\right)}\label{HL1}\\
\frac{d}{dt}b_{n} & = & \mbox{i}\left[H,b_{n}\right]-\gamma_{e}b_{n}+b_{in,n}\label{HL2}\\
\frac{d}{dt}c_{n} & = & \mbox{i}\left[H,c_{n}\right]-\gamma_{r}c_{n}+c_{in,n}\label{HL3}
\end{eqnarray}
where $\left\{ a_{in}^{\left(f\right)},\; a_{in}^{\left(d\right)},\; b_{in,n},\; c_{in,n}\right\} $
denote the respective Langevin forces associated to the incoming fields
from the feeding and detection sides, and to the atomic operators
$b_{n}$ and $c_{n}$. The Hamiltonian of the system $H=H_{0}+H_{int}$
comprises the linearized EIT (later referred to as ``unperturbed'')
Hamiltonian 
\begin{eqnarray}
H_{0} & = & -\Delta_{c}a^{\dagger}a-\sum_{n=1}^{N}\left(\Delta_{e}b_{n}^{\dagger}b_{n}+\Delta_{r}c_{n}^{\dagger}c_{n}\right)\nonumber \\
 & + & \sum_{n=1}^{N}\left(ga^{\dagger}b_{n}+\frac{\Omega_{cf}}{2}c_{n}^{\dagger}b_{n}+\mathrm{H.c.}\right)\label{eq:H_0}
\end{eqnarray}
and the so-called interaction Hamiltonian $H_{int}=H_{dd}+H_{f}$
consisting of the dipole-dipole interaction Hamiltonian $H_{dd}=\sum_{m<n}^{N}\kappa_{mn}c_{m}^{\dagger}c_{n}^{\dagger}c_{m}c_{n}$
and the cavity feeding Hamiltonian $H_{f}=\alpha\left(a+a^{\dagger}\right)$,
where $\kappa_{mn}\equiv C_{6}/\left\Vert \vec{r}_{m}-\vec{r}_{n}\right\Vert ^{6}$.
In the following, we will need the operators expressed in the interaction
picture, \emph{i.e.} the solutions $\left\{ a_{0}\left(t\right),\; b_{n,0}\left(t\right),\; c_{n,0}\left(t\right)\right\} $
of the system Eqs.(\ref{HL1}-\ref{HL3}) in which $H$ is replaced
by $H_{0}$. \vskip 2mm

\paragraph*{Correlation functions in the Schwinger-Keldysh contour formalism.}

\begin{figure}
\centering{}\includegraphics[width=8cm]{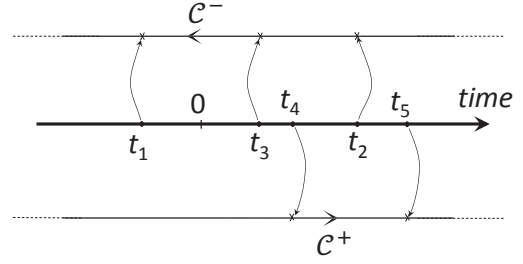} \protect\caption{Representation of contour-ordering for the multitime correlation function
$\left\langle \tilde{\mathcal{T}}\left\{ \mathcal{O}^{1}\left(t_{1}\right)\mathcal{O}^{2}\left(t_{2}\right)\mathcal{O}^{3}\left(t_{3}\right)\right\} \mathcal{T}\left\{ \mathcal{O}^{4}\left(t_{4}\right)\mathcal{O}^{5}\left(t_{5}\right)\right\} \right\rangle $.}
\label{Contour} 
\end{figure}

The correlation function $\left\langle a^{\dagger}\left(t_{1}\right)a\left(t_{2}\right)\right\rangle $,
computed in the vacuum state $\rho_{0}=\left\vert \textrm{Ø}\right\rangle \left\langle \textrm{Ø}\right\vert $,\textcolor{red}{{}
}can be put in the form 
\begin{equation}
\mathrm{Tr}\left\{ \rho_{0}{\cal T}_{{\cal C}}\left(e^{-\mathrm{i}\left(\int_{{\cal C}}H_{int}\right)}a_{0}^{\dagger}\left(t_{1}^{\left(-\right)}\right)a_{0}\left(t_{2}^{\left(+\right)}\right)\right)\right\} \label{Correlation}
\end{equation}
where ${\cal T}_{{\cal C}}$ is the ordering operator along the contour
${\cal C}$ comprising the forward ${\cal C}^{+}\equiv\left(-\infty,\infty\right)$
and backward ${\cal C}^{-}\equiv\left(\infty,-\infty\right)$ parts
\cite{R07,SL13}. ${\cal T}_{{\cal C}}$ is explicitly defined as
follows, for generic Heisenberg operators of the system $\left\{ A_{j}\left(t\right)\right\} $,
\begin{eqnarray*}
 &  & {\cal T}_{{\cal C}}\left(A_{1}\left(t_{1}^{\left(-\right)}\right)\cdots A_{k}\left(t_{k}^{\left(-\right)}\right)A_{k+1}\left(t_{k+1}^{\left(+\right)}\right)\cdots A_{k+l}\left(t_{k+l}^{\left(+\right)}\right)\right)\\
 &  & =\tilde{\mathcal{T}}\left(A_{1}\left(t_{1}^{\left(-\right)}\right)\cdots A_{k}\left(t_{k}^{\left(-\right)}\right)\right)\\
 &  & \times\mathcal{T}\left(A_{k+1}\left(t_{k+1}^{\left(+\right)}\right)\cdots A_{k+l}\left(t_{k+l}^{\left(+\right)}\right)\right)
\end{eqnarray*}
where $\mathcal{T}$, $\tilde{\mathcal{T}}$ denote the usual chronological
and antichronological ordering operators, respectively, and the superscript
$^{\left(\pm\right)}$ indicates the branch $\mathcal{C}^{\pm}$ the
time argument belongs to. More generally, multitime correlation functions
$\left\langle \tilde{{\cal T}}\left\{ \mathcal{O}^{1}\left(t_{1}\right)\cdots\mathcal{O}^{k}\left(t_{k}\right)\right\} {\cal T}\left\{ \mathcal{O}^{k+1}\left(t_{k+1}\right)\cdots\mathcal{O}^{k+l}\left(t_{k+l}\right)\right\} \right\rangle $
involving operators $\mathcal{O}^{j}$ take the form 
\begin{eqnarray}
 &  & \left\langle {\cal T}_{{\cal C}}\left(e^{-\mathrm{i}\left(\int_{{\cal C}}H_{int}\right)}\mathcal{O}_{0}^{1}\left(t_{1}^{\left(+\right)}\right)\cdots\mathcal{O}_{0}^{k}\left(t_{k}^{\left(+\right)}\right)\right.\right.\nonumber \\
 &  & \left.\left.\times\mathcal{O}_{0}^{k+1}\left(t_{k+1}^{\left(-\right)}\right)\cdots\mathcal{O}_{0}^{L}\left(t_{k+l}^{\left(-\right)}\right)\right)\right\rangle \label{eq:O_Plus_O_minus-1}
\end{eqnarray}
where $\mathcal{O}_{0}^{j}\left(t_{j}^{\left(\pm\right)}\right)$
is the operator $\mathcal{O}^{j}$ expressed in the interaction picture.
Fig. \ref{Contour} represents the contour-ordering used for a specific
correlation function. \vskip 2mm

\paragraph*{Perturbative calculation of the correlation functions.}

We now perturbatively expand the correlation function Eq.(\ref{Correlation})
with respect to the feeding Hamiltonian, \emph{i.e.} $\left\langle a^{\dagger}\left(t_{1}\right)a\left(t_{2}\right)\right\rangle =\sum_{k=0}^{+\infty}\left\langle a^{\dagger}\left(t_{1}\right)a\left(t_{2}\right)\right\rangle ^{\left(k\right)}$,
where 
\begin{eqnarray*}
 &  & \left\langle a^{\dagger}\left(t_{1}\right)a\left(t_{2}\right)\right\rangle ^{\left(k\right)}\equiv\frac{\left(-\mathrm{i}\alpha\right)^{k}}{k!}\\
 &  & \times\left\langle {\cal T}_{{\cal C}}\left\{ \left[\int_{{\cal C}}\left(a_{0}+a_{0}^{\dagger}\right)\right]^{k}e^{-\mathrm{i}\left(\int_{{\cal C}}H_{dd}\right)}a_{0}^{\dagger}\left(t_{1}^{\left(-\right)}\right)a_{0}\left(t_{2}^{\left(+\right)}\right)\right\} \right\rangle 
\end{eqnarray*}
and the superscript denotes the order $k$ in the feeding rate $\alpha$.
In the expansion, only remain the terms which contain the same number
of creation and annihilation operators, corresponding to even $k$'s.
These terms can be further evaluated by perturbatively expanding them
with respect to the dipole-dipole interaction Hamiltonian $H_{dd}$.
From now on, we shall more specifically be interested in the temporal
Fourier transform of the correlation function $\left\langle a^{\dagger}\left(\omega_{1}\right)a\left(\omega_{2}\right)\right\rangle $.
Following Wick's theorem \cite{R07,SL13}, the contribution of each
order in $H_{dd}$ is a sum of products of the Fourier transformed
time-ordered non-perturbed Green's functions defined by

\begin{eqnarray}
G_{x_{0}y_{0}}\left[\omega\right] & \equiv & -{\rm i}\int dte^{i\omega t}\left\langle {\cal T}\left(x_{0}\left(t\right)y_{0}^{\dagger}\left(0\right)\right)\right\rangle \label{eq:Greens}
\end{eqnarray}
where $\left(x_{0},y_{0}\right)$ stand for any 2 bosonic operators
of the system in the interaction picture. These functions may be readily
computed from the (linear) Bloch equations spanned by $H_{0}$.

The full resummation of the perturbation series in $H_{dd}$ can be
performed for the first four orders in $\alpha$. For the second order,
we get 
\begin{eqnarray}
\left\langle a^{\dagger}\left(\omega_{1}\right)a\left(\omega_{2}\right)\right\rangle ^{\left(2\right)} & = & \left\langle a^{\dagger}\left(\omega_{1}\right)\right\rangle ^{\left(1\right)}\left\langle a\left(\omega_{2}\right)\right\rangle ^{\left(1\right)}\label{eq:SecondOrder}
\end{eqnarray}
where $\left\langle a\left(\omega\right)\right\rangle ^{\left(1\right)}=\sqrt{2\pi}\alpha\delta\left(\omega\right)G_{a_{0}a_{0}}\left[0\right]$
is the first order average value for the cavity mode annihilation
operator. Note that Eq.(\ref{eq:SecondOrder}) exhibits the factorization
property we pointed out and used in a previous work \cite{GBB15}.
By contrast, the fourth order $\left\langle a^{\dagger}\left(\omega_{1}\right)a\left(\omega_{2}\right)\right\rangle ^{\left(4\right)}$
contains a factorized part 
\begin{eqnarray}
 &  & \left\langle a^{\dagger}\left(\omega_{1}\right)\right\rangle ^{\left(1\right)}\left\langle a\left(\omega_{2}\right)\right\rangle ^{\left(3\right)}+\left\langle a^{\dagger}\left(\omega_{1}\right)\right\rangle ^{\left(3\right)}\left\langle a\left(\omega_{2}\right)\right\rangle ^{\left(1\right)}\label{eq:Expan_4orderEl}
\end{eqnarray}
but also an extra component, which shall be denoted by $\left\langle a^{\dagger}\left(\omega_{1}\right),a\left(\omega_{2}\right)\right\rangle ^{\left(4\right)}$
-- this term is actually a covariance and characterizes the interaction-induced
non-classicality of the system. More explicitly, the calculations
yield 
\begin{eqnarray}
\left\langle a\left(\omega\right)\right\rangle ^{\left(3\right)} & = & \delta\left(\omega\right)\sqrt{2\pi}\alpha^{3}\left(G_{s_{0}a_{0}}\left[0\right]\right)^{2}T_{0}\left|G_{a_{0}s_{0}}\left[0\right]\right|^{2}\label{Elastic}
\end{eqnarray}
and 
\begin{eqnarray}
 &  & \left\langle a^{\dagger}\left(\omega_{1}\right),a\left(\omega_{2}\right)\right\rangle ^{\left(4\right)}\nonumber \\
 & = & -2\alpha^{4}\delta\left(\omega_{1}-\omega_{2}\right)\left|G_{a_{0}s_{0}}\left[\omega_{1}\right]\right|^{2}\left|T_{0}\right|^{2}\label{Inelastic}\\
 &  & \times\left|G_{s_{0}a_{0}}\left[0\right]\right|^{4}\Im\left(G_{s_{0}s_{0}}\left[-\omega_{1}\right]\right)\nonumber 
\end{eqnarray}
The previous expressions involve Green's functions relative to the
symmetric Rydberg spinwave annihilation operator $s_{0}\equiv\frac{1}{\sqrt{N}}\sum_{n}c_{n,0}$
as well as the term $T_{0}$ which accounts for the quantum interference
of all possible scattering processes of two symmetric Rydberg spinwaves
at all orders in $H_{dd}$. The value of $T_{0}$ takes into account
the resummation of the series in $H_{dd}$, and it can be evaluated
as the specific element $T_{0}\equiv T\left[0,0\right]$ of the matrix
$T\left[\vec{k},\vec{k}'\right]$ which describes the scattering of
two incoming into two outgoing Rydberg spinwaves of respective wavevectors
$\left(\vec{k},-\vec{k}\right)$ and $\left(\vec{k}',-\vec{k}'\right)$,
as considered in \cite{BCF14}. Using, e.g., diagrammatic techniques
\cite{AGD}, one derives the self-consistent equation 
\begin{equation}
T\left[\vec{k},\vec{k}^{\prime}\right]=U_{\vec{k}-\vec{k}^{\prime}}+\mbox{i}\sum_{\vec{q}}S_{\vec{q}}U_{\vec{k}-\vec{q}}T\left[\vec{q},\vec{k}^{\prime}\right]\label{selfconsistent}
\end{equation}
where $S_{\vec{q}}\equiv\frac{1}{2\pi}\int\mbox{d}\omega\: G_{s_{-\vec{q}}s_{-\vec{q}}}\left[-\omega\right]G_{s_{\vec{q}}s_{\vec{q}}}\left[\omega\right]$,
$s_{\vec{q}}\equiv\frac{1}{\sqrt{N}}\sum_{n}e^{-\mbox{i}\vec{q}.\vec{r}_{n}}c_{n,0}$
is the Rydberg spinwave with the wavevector $\vec{q}$ and $U_{\vec{K}}$
is the spatial Fourier transform of the interaction potential $C_{6}/\left\Vert \vec{r}\right\Vert ^{6}$.
It can be shown that for $\vec{q}\neq0$ all Green's functions $G_{s_{\vec{q}}s_{\vec{q}}}\left[\omega\right]$
coincide whence $S_{\vec{q}\neq0}\equiv S$ and therefore Eq. (\ref{selfconsistent})
becomes 
\begin{equation}
T\left[\vec{k},\vec{k}^{\prime}\right]=U_{\vec{k}-\vec{k}^{\prime}}+\mbox{i}S_{0}U_{\vec{k}}T\left[0,\vec{k}^{\prime}\right]+\mbox{i}S\sum_{\vec{q}\neq0}U_{\vec{k}-\vec{q}}T\left[\vec{q},\vec{k}^{\prime}\right]\label{selfconsistentbis}
\end{equation}
$T$ can now be related to its value, denoted $\tilde{T}$ in the
fictitious situation when $g=0$, \emph{i.e.} atoms are decoupled
from the cavity. In that case, one indeed shows that $S_{0}=S$ and
therefore Eq. (\ref{selfconsistentbis}) yields 
\begin{equation}
\tilde{T}\left[\vec{k},\vec{k}^{\prime}\right]=U_{\vec{k}-\vec{k}^{\prime}}+\mbox{i}S\sum_{\vec{q}}U_{\vec{k}-\vec{q}}\tilde{T}\left[\vec{q},\vec{k}^{\prime}\right]\label{selfconsistentter}
\end{equation}
Finally, from Eqs.(\ref{selfconsistentbis},\ref{selfconsistentter}),
one gets

\begin{equation}
T\left[\vec{k},\vec{k}^{\prime}\right]=\tilde{T}\left[\vec{k},\vec{k}^{\prime}\right]+\frac{\mbox{i}\left(S_{0}-S\right)\tilde{T}\left[\vec{k},0\right]\tilde{T}\left[0,\vec{k}^{\prime}\right]}{1-\mbox{i}\left(S_{0}-S\right)\tilde{T}\left[0,0\right]}\label{eq:Tkk}
\end{equation}
and therefore $T_{0}=\tilde{T}_{0}/\left[1-\mbox{i}\left(S_{0}-S\right)\tilde{T}_{0}\right]$
where we defined $\tilde{T}_{0}\equiv\tilde{T}\left[0,0\right]$.
When $g=0$, Rydberg excitations cannot hop from the atom where they
were created to another : the calculation of $\tilde{T}_{0}$ is therefore
a simple two-body problem and, for a sample of volume $V$, we find
\[
\tilde{T}_{0}=\frac{1}{N^{2}}\sum_{mn}\frac{\kappa_{mn}}{1-\mathrm{i}\kappa_{mn}S}\approx-\frac{2\pi^{2}}{3V}\sqrt{\frac{-{\rm i}\left|C_{6}\right|}{S}}
\]
In principle, we can numerically compute physical quantities for arbitrary
excitation numbers $\left(k\geq2\right)$, provided that excitation
exchange between atoms via the cavity mode is neglected; this approximation
is valid for samples much larger than the blockade radius $\left(v_{b}\ll V\right)$.
The calculations involve the so-called Faddeev equations and will
be presented elsewhere. In the following section we present and analyze
an exact analytic expression derived from the two-excitation computation
above. \vskip 2mm

\paragraph*{Transmitted light spectrum.}

\begin{figure}
\begin{centering}
\includegraphics[width=7cm]{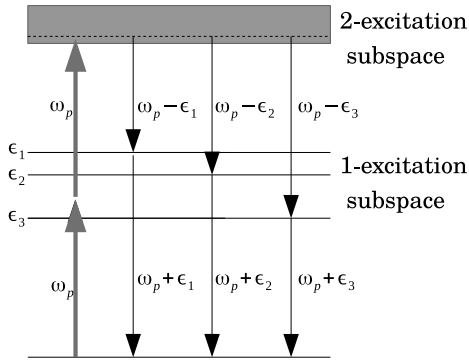} 
\par\end{centering}

\protect\caption{The level structure of the Hamiltonian (\ref{eq:H_0}) restricted
to two excitations. The structure of the doubly excited manifold is
represented schematically.}

\label{Polariton} 
\end{figure}

\begin{figure}
\begin{raggedright} \emph{\Large{}{}a}{\Large{}{})} 

\end{raggedright}

\begin{centering}
\includegraphics[scale=0.77]{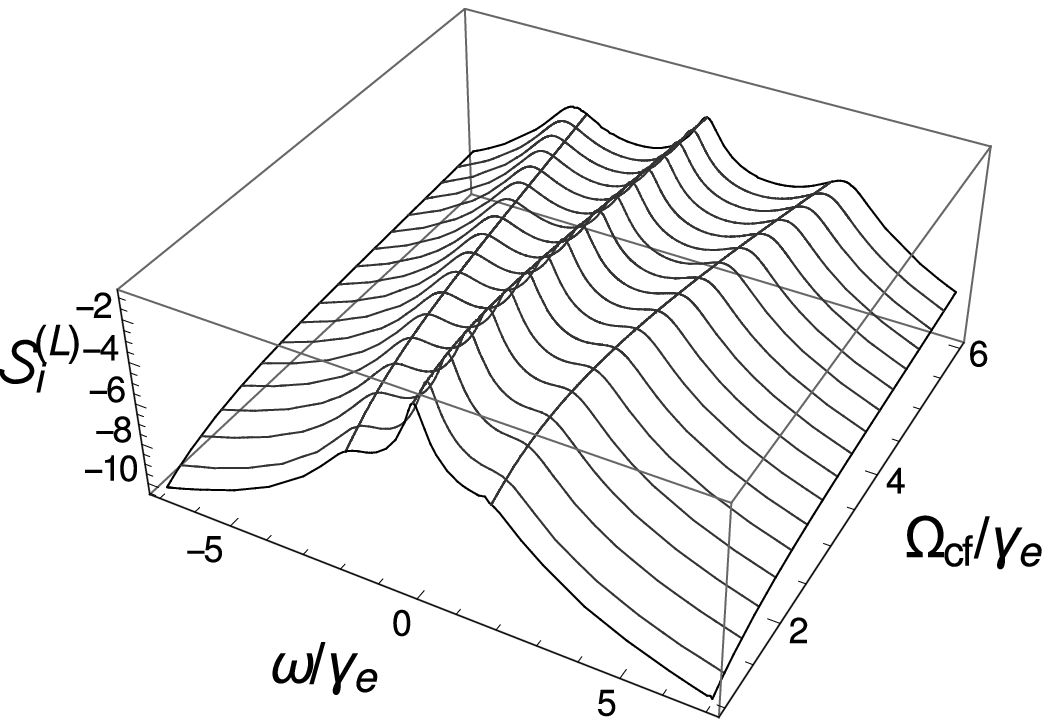} 
\par\end{centering}

\begin{raggedright} \emph{\Large{}{}b}{\Large{}{})} 

\end{raggedright}

\begin{centering}
\includegraphics[scale=0.77]{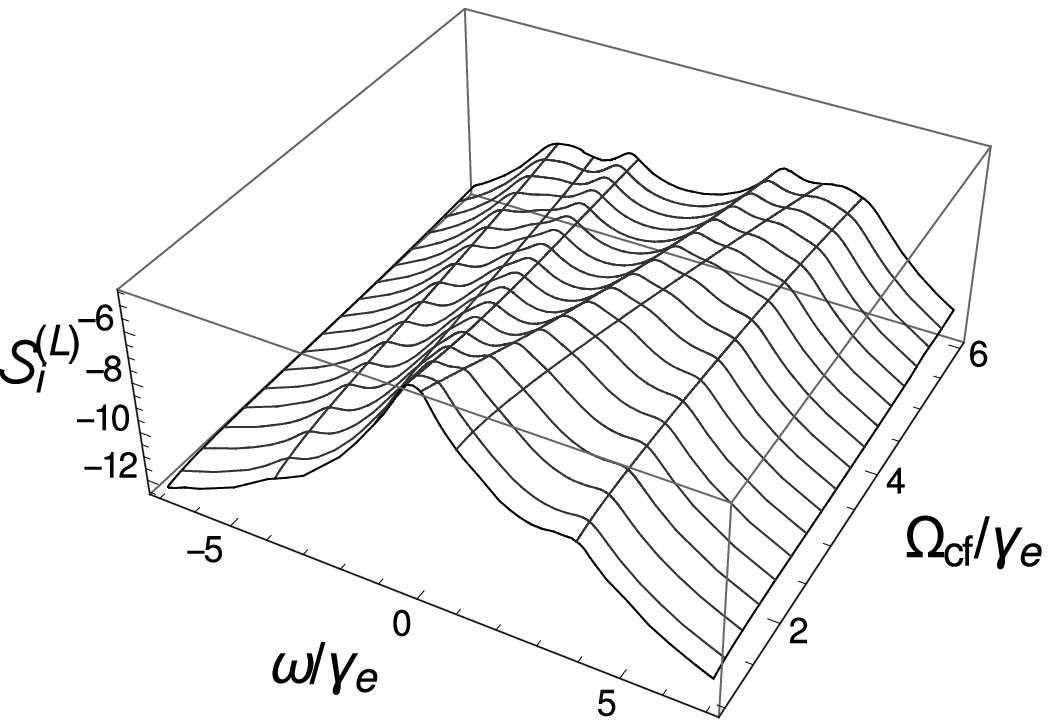} 
\par\end{centering}

\protect\caption{Logarithm of the cavity transmission spectrum ${\cal S}_{i}^{\left(L\right)}\equiv\log_{10}\left[2\gamma_{c}^{\left(d\right)}\int d\nu\left\langle a^{\dagger}\left(\omega\right),a\left(\nu\right)\right\rangle ^{\left(4\right)}\right]$
at fourth order as a function of $\Omega_{cf}$ and the frequency
(in the frame rotating at $\omega_{p}$) for : a) the resonant case
$\Delta_{c}=\Delta_{e}=\Delta_{r}=0$, b) the detuned case. The transverse
curves give $\left(\pm\epsilon_{1},\pm\epsilon_{2},\pm\epsilon_{3}\right)$
as functions of $\Omega_{cf}$ (see main text).}

\label{Spectrum} 
\end{figure}

The previous results allow us to investigate the spectrum of the light
transmitted through the cavity, \emph{i.e.} $\mathcal{S}\left(\omega\right)\equiv\int\mbox{d}\nu\:\left\langle a_{out}^{\left(d\right)\dagger}\left(\omega\right)a_{out}^{\left(d\right)}\left(\nu\right)\right\rangle =2\gamma_{c}^{\left(d\right)}\int\mbox{d}\nu\:\left\langle a^{\dagger}\left(\omega\right)a\left(\nu\right)\right\rangle $
\cite{WM08}. Expanding $\mathcal{S}\left(\omega\right)$ up to the
fourth order in $\alpha$ and using Eqs.(\ref{eq:Expan_4orderEl}-\ref{Inelastic}),
we find $\mathcal{S}\left(\omega\right)\approx\mathcal{S}^{\left(2\right)}\left(\omega\right)+\mathcal{S}^{\left(4\right)}\left(\omega\right)$,
where $\mathcal{S}^{\left(2\right)}\left(\omega\right)=4\gamma_{c}^{\left(d\right)}\pi\alpha^{2}\left|G_{a_{0}a_{0}}\left[0\right]\right|^{2}\delta\left(\omega\right)$
while $\mathcal{S}^{\left(4\right)}\left(\omega\right)=\mathcal{S}_{e}^{\left(4\right)}\left(\omega\right)+\mathcal{S}_{i}^{\left(4\right)}\left(\omega\right)$
with 
\begin{eqnarray*}
\mathcal{S}_{e}^{\left(4\right)}\left(\omega\right) & = & 4\gamma_{c}^{\left(d\right)}\Re\left[\int\mbox{d}\nu\:\left\langle a^{\dagger}\left(\omega\right)\right\rangle ^{\left(1\right)}\left\langle a\left(\nu\right)\right\rangle ^{\left(3\right)}\right]\propto\delta\left(\omega\right)\\
\mathcal{S}_{i}^{\left(4\right)}\left(\omega\right) & = & 2\gamma_{c}^{\left(d\right)}\int\mbox{d}\nu\:\left\langle a^{\dagger}\left(\omega\right),a\left(\nu\right)\right\rangle ^{\left(4\right)}
\end{eqnarray*}
In other words, whereas at the second order in $\alpha$ the transmitted
light has the same frequency as the probe field, the fourth order
spectrum comprises both an elastic contribution $\mathcal{S}_{e}^{\left(4\right)}\left(\omega\right)$
-- which partially compensates the second order contribution $\mathcal{S}^{\left(2\right)}\left(\omega\right)$,
and an inelastic contribution $\mathcal{S}_{i}^{\left(4\right)}\left(\omega\right)$
which renders the transmitted light slightly polychromatic. Though
physically expectable, this behaviour had never been reported, to
our knowledge. Fig. \ref{Spectrum} displays the fourth-order inelastic
component $\mathcal{S}_{i}^{\left(4\right)}$ as a function of the
frequency $\omega$ and the control field Rabi frequency $\Omega_{cf}$
in: (a) resonant ($\Delta_{c}=\Delta_{e}=\Delta_{r}=0$) and (b) detuned
($\Delta_{c}=-3\gamma_{e},\Delta_{e}=0,\Delta_{r}=0$) configurations.
We moreover assume a cloud cooperativity $C\equiv\frac{g^{2}N}{2\gamma_{c}\gamma_{e}}=5$,
and $\gamma_{c}^{\left(d\right)}=0.3\gamma_{e}\gg\gamma_{c}^{\left(f\right)}$
, $\gamma_{r}=0.15\gamma_{e}$ and $\gamma_{e}=2\pi\times3{\rm MHz}$
for the cavity, Rydberg and intermediate state decays, respectively.
In both cases, $\mathcal{S}_{i}^{\left(4\right)}\left(\omega\right)$
shows resonances -- three for the resonant configuration (Fig.\ref{Spectrum}
a), six for the generic detuned case (Fig.\ref{Spectrum} b), whose
frequencies' absolute values correspond to the three polariton eigenenergies
$\left\{ \epsilon_{k=1,2,3}\right\} $ of the Hamiltonian in the single
excitation subspace, written in the frame rotating at the probe frequency
\begin{equation}
\left(\begin{array}{ccc}
-\Delta_{c} & g\sqrt{N} & 0\\
g\sqrt{N} & -\Delta_{e} & \frac{\Omega_{cf}}{2}\\
0 & \frac{\Omega_{cf}}{2} & -\Delta_{r}
\end{array}\right)\label{eq:Single_Ham}
\end{equation}
This can be understood by noticing that $\mathcal{S}_{i}^{\left(4\right)}\left(\omega\right)$
corresponds to a four-photon process, in which two probe photons are
absorbed by the system which subsequently reemits two photons through
radiative or cavity decays via the three polariton states (see Fig.
\ref{Polariton}) \cite{OKK}. Three two-photon emission channels
are possible, corresponding to the three polariton energies $\epsilon_{k}$.
For each channel $\left(k\right)$, the reemitted photons' frequencies
are respectively given by $\left(\omega_{p}\pm\epsilon_{k}\right)$,
that is $\pm\epsilon_{k}$ in the frame rotating at the probe frequency,
which leads up to six resonances, seen in Fig. \ref{Spectrum} b).
In the specific resonant case (a), $\epsilon_{1}=\epsilon_{3}$ and
therefore only three resonances can be identified (cf Fig. \ref{Spectrum}
a). Let us note that, since $H_{dd}$ enters Eq.(\ref{Inelastic})
only via the overall factor $T_{0}$, the strength of dipole-dipole
interactions does not affect the $\omega$-dependence of the inelastic
component at fourth order but merely governs its order of magnitude.
\vskip 2mm

\paragraph*{Conclusion.}

In this Letter, we investigated the quantum optical nonlinearities
induced by a cavity Rydberg EIT medium. Using the Schwinger-Keldysh
contour formalism, we analytically computed field correlation functions
beyond the lowest non-vanishing order in feeding, and the transmitted
light spectrum. Its inelastic part appears to contain several resonances
explained by a simple polaritonic picture. For simplicity, we assumed
a constant feeding of the system: the formalism allows, however, for
time-dependent wavepacket inputs as well. Though rarely employed in
this context, the contour formalism is a powerful tool for quantum
optics which could be used, e.g., to compute higher-order correlation
functions of the system or thoroughly analyze subtle effects in Rydberg
atomic ensembles such as thermalization \cite{AGL12} or phase transition
\cite{LWK09}. 
\begin{acknowledgments}
This work is supported by the European Union grants SIQS (FET \#600645)
and RySQ ( FET \# 640378), and by the « Chaire SAFRAN - IOGS Photonique
Ultime ». 
\end{acknowledgments}

$^{\dagger}$ Present address: Collège de France, 11 Place Marcelin
Berthelot, 75005 Paris, France.

\end{document}